# Privacidade digital como direito do cidadão: o caso dos grupos indígenas do Brasil*

**Yasodara Córdova (yasodara.cordova@hks.harvard.edu)**
*Digital Harvard Digital Kennedy School*

*O PRESENTE ARTIGO É UM RASCUNHO

**Resumo**

Grupos de indígenas brasileiros, podem ser profundamente afetados pelo abuso no uso de seus dados. Nesse sentido, questões que envolvem o uso de dados pessoais de indígenas brasileiros explodem de vez em quando, muitas vezes sem resposta institucional adequada, exceto para casos isolados - como quando as amostras de sangue da tribo Yanomami, consideradas valiosas pela diversidade de sua genética, retornaram às suas proprietários em consequência de uma decisão judicial, após quase trinta anos de espera. No entanto, existem casos em que os dados sensíveis de tribos inteiras continuam a fluir por caminhos mal conhecidos e desprotegidos, colocando em risco seu direito à privacidade e soberania.

A exploração arbitrária da diversidade genética, cultural e dos corpos dos povos indígenas no Brasil desdenha qualquer separação entre o caráter pessoal e privado da informação e dos interesses dos mesmos, potencialmente relacionados também à propriedade intelectual, muitas vezes é suportada por regras fracas e plataformas que trazem na coleta de dados distorções a serem consolidadas em algoritmos em futuro próximo. Existe ainda o caso de comunidades inteiras que seguem sem contato por escolha própria, tendo sua privacidade precária salva dos predadores digitais apenas pela barreira da falta de conectividade da floresta. Este relatório pretende demonstrar a situação da proteção de dados sensíveis de pessoas indígenas no Brasil, desde o seu quadro regulatório até os sistemas utilizados para armazená-lo e extraí-lo e distribuí-lo, ao mesmo tempo que nomeia medidas e tópicos a serem considerados na discussão.

# INTRODUCÃO

*"We create a society that has no use for the disabled or the elderly, and then are cast aside when we are hurt or grow old. We measure human worth based only on the ability to earn a wage and suffer in a world that undervalues care and community.*

*We base our economy on exploiting the labor of racial and ethnic minorities, and watch lasting inequities snuff out human potential. We see the world as inevitably riven by bloody competition and are left unable to recognize the many ways we cooperate and lift each other up"*

*— Virginia Eubanks*[1]

O processo de digitalização dos corpos através da coleta de dados surge em momento que pode levar tanto à inovação em práticas de prestação de serviços para cidadãos, quanto à solidificação das desigualdades, a partir de um processo de vigilância progressivo. De um jeito ou de outro, "O futuro já está aqui - só não é distribuído de maneira uniforme". O estudo sobre os sistemas que servem de plataforma para a coleta das informações pessoais dos indígenas brasileiros oferece pistas sobre como as premissas utilizadas para a criação desses sistemas podem partir de paradigmas colonizadores, com a solidificação de um sistema de vigilância destinado a apagar vestígios de culturas diversificadas também no mundo digital.

Muitas comunidades originais, degradadas pelo processo de colonização, perderam parte de sua cultura e resistência por métodos de substituição de identidade. Durante a invasão portuguesa, a prática do casamento forçado de estrangeiros com jovens mulheres indígenas no Brasil, por exemplo, bem como as práticas de batismo forçado foram destinadas a reforçar a ideia de que não havia nação civilizada naquela terra antes, portanto os indígenas faziam parte da natureza não tendo direito à cidadania, que era mantida como privilégio dos colonizadores.

Essas são práticas que incorporaram o etos da colonização na América do Sul, "visando à destruição, submissão e exploração das sociedades do Novo Mundo" (Salles, 1996, p. 83, 98-9) acabaram por ratificar a violência, tanto física quanto simbólica dos povos originários. A alienação do povo de suas identidades originais lhes negou o direito à dignidade e autonomia sobre sua terra, corpo, cultura e práticas religiosas.

Mais de quatro séculos depois do desembarque de Pedro Álvares Cabral no Brasil (1500) e os indígenas brasileiros adquiriram o direito de ter uma identidade oficial emitida pelo Estado brasileiro; daí sua cidadania, concedida pela Constituição brasileira em 1988. A determinação da capacidade dos povos indígenas depende de uma série de resoluções interligadas. No passado, foi exigida a tutela do indígena para que exercessem atos civis. Embora existam no congresso nacional projetos de lei que pretendem regular a participação das comunidades indígenas como cidadãos com plena capacidade, estes permanecem paralisados, com pouca perspetiva de discussão nos próximos anos. Os sistemas usados para rastrear e identificar essas comunidades estão inseridos em um ecossistema que traz o paternalismo populista como paradigma da integração dos indígenas como cidadãos brasileiros. Desse modo, para exercer a tutela sob os indígenas, estabeleceu-se a criação de órgãos especiais, responsáveis inclusive pelo registro de nascimento com o reconhecimento de etnias:

> “A legislação especial dispõe que os índios e as comunidades indígenas ainda não integrados à comunhão nacional ficam sujeitos ao regime tutelar da União que será exercido por meio da Fundação Nacional do Índio - FUNAI, fundação pública vinculada ao Ministério da Justiça, criada pela Lei nº 5.371, de 5 de dezembro de 1967, em substituição ao antigo Serviço de Proteção ao Índio que datava de 1910. Nos termos do Estatuto do Índio, "são considerados nulos os atos praticados entre índios não integrados e qualquer pessoa estranha à comunidade indígena quando não tenha havido assistência do órgão tutelar competente" (art. 7.º, §8.º, da Lei nº 6.001/73).[2]

[1] Eubanks, V., 2018. Automating inequality: How high-tech tools profile, police, and punish the poor. St. Martin's Press.

Ainda que a condição de indígenas seja autodeclarada, estabelece-se que os indígenas que não estejam integrados possam ser tutelados caso não tenham entendimento sobre os procedimentos envolvidos no exercício da cidadania. Com a digitalização de processos e das identidades do cidadão, o processo tende a se agravar. Este princípio implica não só na exclusão da possibilidade de que as práticas indígenas sejam incorporadas ao fazer do Estado brasileiro, mas também na vigilância e submissão dos povos indígenas aos sistemas que perpetuam o etos colonizador, sendo também impostos por registros e bancos de dados de modo ainda mais duradouro.

## 1) Quais dados?

Dados sobre indígenas brasileiros são facilmente encontrados na web. Há várias instituições que cuidam de estabelecer as pesquisas e manter atualizadas as bases de governo sobre populações originais do Brasil, não sendo difícil fazer download de pastas inteiras com localização geográfica de comunidades, muitas vezes compostas por poucos indivíduos, de idades diferentes, de fácil deanonimização. Essa distribuição de informações sobre indígenas via Internet tem duas causas: a necessidade do Estado de vigiar o desenvolvimento das comunidades, perante o argumento de que o monitoramento traria melhoria de políticas públicas e proteção às mesmas; e o cumprimento de tratados internacionais referentes à conservação dos povos originais.

[2] Aspectos atuais da capacidade dos índios, Átila Da Rold Roesler, setembro de 2010. Disponível em: https://jus.com.br/artigos/17328/aspectos-atuais-da-capacidade-civil-dos-indios

É uma preocupação mundial que os governos de países que abrigam povos indígenas possam demonstrar a habilidade para cuidar de suas populações originais. Para garantir internacionalmente que está a cumprir os tratados, o Governo Brasileiro tem optado por se concentrar em ações de coleta e distribuição de dados, que sozinhas, sem o devido acompanhamento por melhorias nas políticas públicas, tornam-se perniciosas à própria população indígena. Dentre as plataformas que abrigam e possibilitam a coleta de dados de indígenas brasileiros as mais importantes são as que se relacionam com dados da saúde indígena, portanto dados pessoais e de caráter extremamente delicado. Os esforços de coleta desses dados no Brasil se explicam por meio de intrincada rede de responsabilidades, estabelecida pela própria legislação.

O Subsistema de Atenção à Saúde dos Povos Indígenas foi criado em 1999, por meio da Lei nº 9.836/99, junto com os Distritos Sanitários Especiais Indígenas (DSEIs). Tais distritos compõem os pólo-base, hoje unidades que funcionam de modo assíncrono em termos de envios e coleta de dados nos locais onde existem reservas indígenas demarcadas ou em processo de demarcação. As DSEIs, além de terem autonomia para coleta e gestão de dados, ainda exercem outras funções dentro dos sistemas de saúde, tal como distribuição e armazenamento de medicamentos, investigação epidemiológica, etc. Como existem problemas de conexão à internet, as DSEIs trabalham de modo assíncrono, com autonomia para coletar os dados, mas sem autonomia para definir quais os principais parâmetros na construção dos sistemas, na definição sobre quais dados serão ou não preservados, transmitidos ou protegidos. Os DSEIs são pontos de agregação de dados, assim como Aldeia, Etnia, Município, Estado, Regiões de Saúde, Região do Brasil e Etnia. As DSEIs estão espalhadas pelo Brasil, e sua localização está disposta na Página do Ministério da Saúde[3], uma a uma. Um exemplo é a imagem a seguir, que detalha a localização das áreas de atuação da DSEI Yanomami.

---

[3] Distritos Sanitários Especiais Indígenas (DSEIs) Accessado em 17 de setembro de 2018. Accessível em http://portalms.saude.gov.br/saude-indigena/saneamento-e-edificacoes/dseis

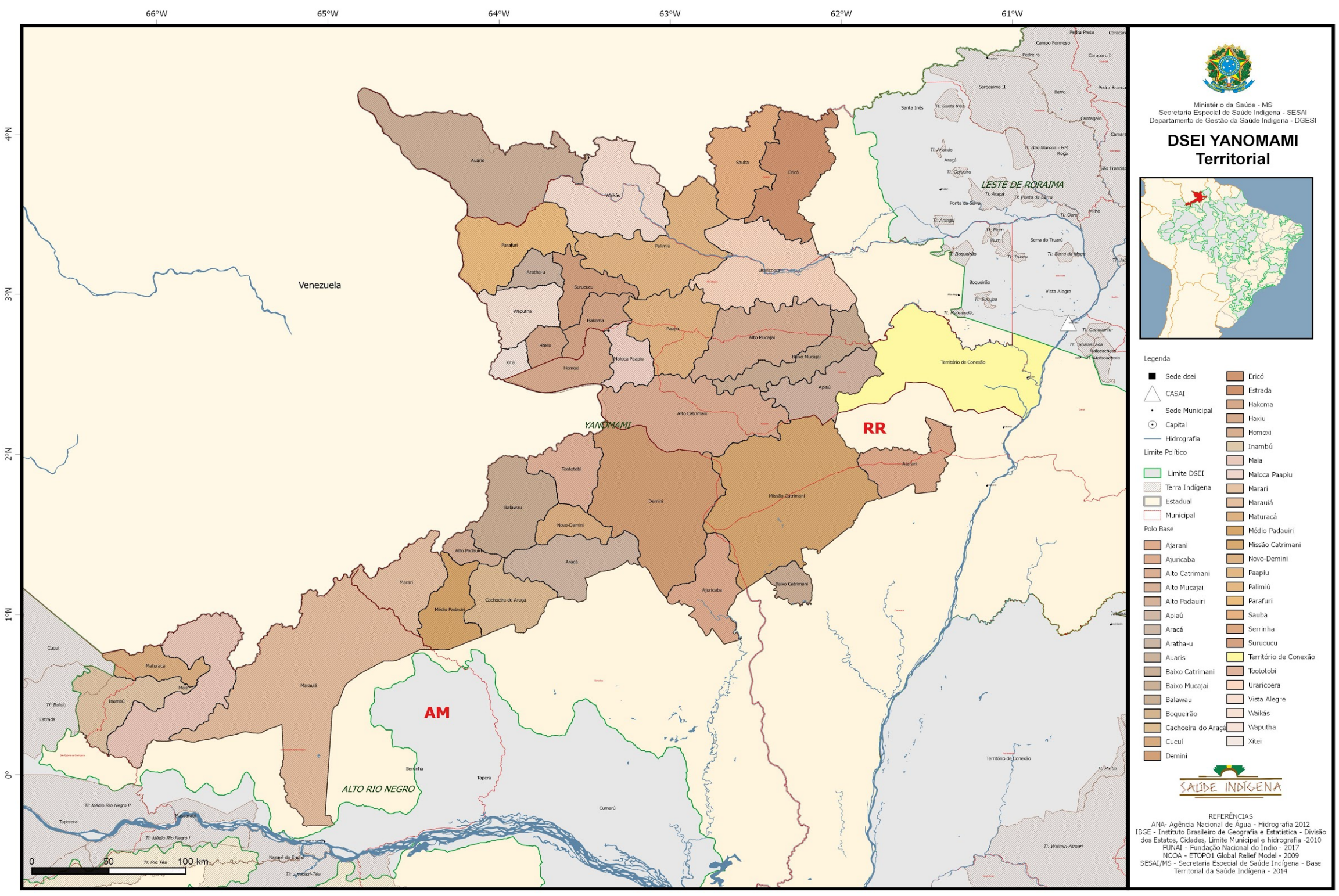

*Figura 1. Imagem de [DSEI Yanomami]*

Como consequência da criação do subsistema de saúde indígena se deu a criação do Sistema de Informação da Atenção à Saúde Indígena, conhecido como SIASI. Este sistema consiste em uma série de softwares voltados para a criação de inteligência em torno de dados e informações relativos à saúde indígena, mas que também prevê a integração com outros sistemas de outras autarquias. Mencionado na portaria do Ministério da Saúde de n. 254, de 6 de Fevereiro de 2002 [4], o SIASI ficou sob os cuidados da Fundação Nacional de Saúde[5] (FUNASA), entidade do Ministério da Saúde também responsável por promover o saneamento básico da população, até 2010. A partir de então o SIASI ficou sob responsabilidade conjunta do núcleo de Tecnologia da Informação (NTI/DAB/SAS/Ministério da Saúde) em conjunto com o DATASUS.

Outras informações que seriam utilizadas para a melhoria de políticas públicas para os indígenas são provenientes de outros sistemas, todos ligados à vigilância em saúde. São eles o 1) Sistema Nacional de Notificação de Agravos (*Sinan*); 2) Sistema de informações de Mortalidade (Sim); 3) Sistema de Vigilância Alimentar e Nutricional (Sisvan); e o 4) Sistema de informações de Vigilância Epidemiológica (Sivep) da Secretaria de Vigilância em Saúde do Ministério da Saúde.

[4] Políticas públicas para de atencão à saúde indígena: documento base. Portaria do Ministério da Saúde de n. 254, de 6 de Fevereiro de 2002, acessada em: 12 de novembro de 2017. Disponível em: http://bvsms.saude.gov.br/bvs/publicacoes/politica_saude_indigena.pdf

[5] Portal da Fundação Nacional de Saúde (Funasa). Acessado em 10 de dezembro de 2017. Disponível em: http://www.funasa.gov.br/competencias

Outros indicadores deveriam ser incorporados ao estudo desses sistemas, mas a comunicação dos dados provenientes das DSEIs não prevê dupla via no envio de informações, o que poderia garantir a descentralização da gestão e a boa formulação de políticas públicas localmente. Também a participação da própria comunidade indígena na gestão de seus dados e definição de paradigmas para estabelecer parâmetros de coleta seria importante para evitar o prejuízo das mesmas. Sabe-se também que as causas da péssima situação da saúde indígena no Brasil estão também ligadas à situação precária onde se encontram as comunidades, que, sem suas terras demarcadas e sofrendo a pressão do agronegócio, encontram-se em situação muito vulnerável. A gestão desses dados, porém, apesar de contar com a coleta assíncrona, é centralizada. A Sesai, a Secretaria especial de saúde indígena, "responsável por coordenar e executar a Política Nacional de Atenção à Saúde dos Povos Indígenas e todo o processo de gestão do Subsistema de Atenção à Saúde Indígena (SasiSUS) no Sistema Único de Saúde (SUS)[6]" concentra as decisões relativas aos sistemas e à sua integração com outras bases.

Os dados que compõem essas bases de dados do SIASI são, em sua maioria, pessoais e deveriam ser considerados íntimos. A concepção da arquitetura da gestão evidencia a falta de agência dos próprios indígenas, até pela inexistência de legislação específica para proteção da sua privacidade. Listados na tabela seguinte se encontram alguns dos parâmetros para coleta dos dados de comunidades indígenas.

Tabela 1

Estruturação e funcionamento do Sistema de Informação da Atenção à Saúde Indígena (SIASI).

| Módulo | Finalidade | Principais variáveis | Implantação/Funcionamento |
|---|---|---|---|
| Demográfico | Informações demográficas/ cadastro individual | Parentesco, nascimento, óbito, residência | Implantado em 2000, com as modalidades local e Web |
| Morbidade | Informações epidemiológicas e de produtividade das unidades e dos profissionais | Recursos humanos, procedimentos, diagnósticos, exames laboratoriais | Em fase de implantação na modalidade local. Funcionando na modalidade Web |
| Imunização | Informações de vacinas aplicadas e planejamento de doses a aplicar | Idade, vacinas aplicadas e a aplicar | Em fase de implantação na modalidade local. Funcionando na modalidade Web |
| Nutrição | Informações sobre o estado nutricional dos grupos de alto risco (idosos, crianças, gestantes e nutrizes) | * | Não implantado. Para funcionamento na modalidade local |
| Saúde bucal | Informações sobre as ações de saúde bucal | * | Não implantado. Para funcionamento na modalidade local |
| Acompanhamento à gestação | Informações sobre o pré-natal | * | Não implantado. Para funcionamento na modalidade local |
| Recursos humanos | Informações de capacitação, formação e quantitativo de pessoal | * | Não implantado. Para funcionamento na modalidade Web |
| Infra-estrutura | Informações sobre infra-estrutura | * | Não implantado. Para funcionamento na modalidade Web |

* Ainda não desenvolvido.
Fonte: Fundação Nacional de Saúde [24].

*Figura 2. tabela SIASI*

[6] Portal da Secretaria Especial de Saúde Indígena (Sesai). Acessado em 12 de setembro de 2018. Disponível em: http://portalms.saude.gov.br/sesai

O descaso com a privacidade dos indivíduos indígenas fica evidente na exibição de dados sensíveis de localização, além de informações como quantos membros existem por etnia, o sexo, e até idade, como se pode ver nas imagens abaixo, retiradas de sites do Ministério da Saúde que fornecem relatórios sobre os indígenas, sob acesso livre.

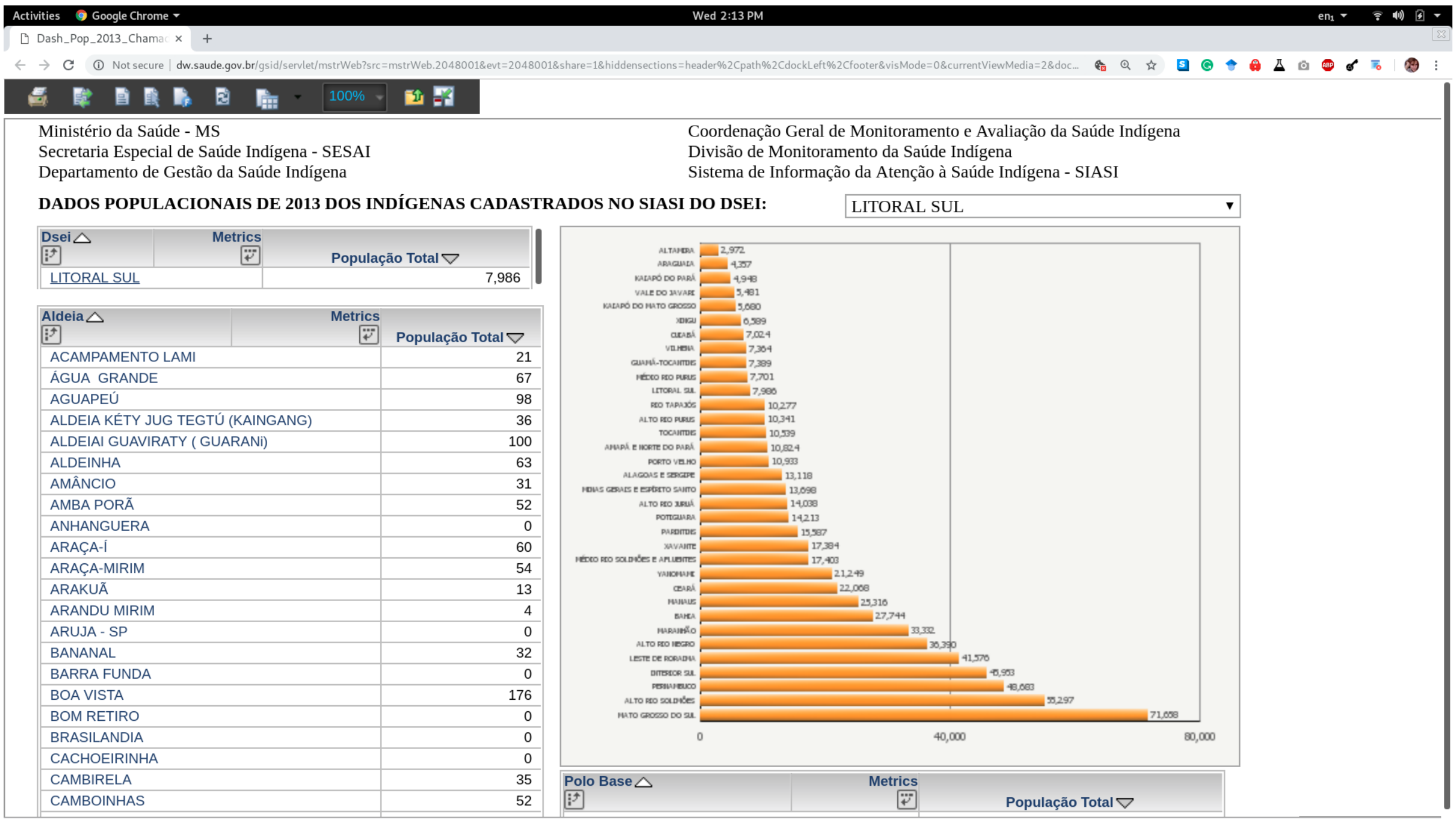

| Dsei | Metrics População Total |
|---|---|
| LITORAL SUL | 7,986 |

| Aldeia | Metrics População Total |
|---|---|
| ACAMPAMENTO LAMI | 21 |
| ÁGUA GRANDE | 67 |
| AGUAPEÚ | 98 |
| ALDEIA KÉTY JUG TEGTÚ (KAINGANG) | 36 |
| ALDEIAI GUAVIRATY ( GUARANI) | 100 |
| ALDEINHA | 63 |
| AMÂNCIO | 31 |
| AMBA PORÃ | 52 |
| ANHANGUERA | 0 |
| ARAÇA-Í | 60 |
| ARAÇA-MIRIM | 54 |
| ARAKUÃ | 13 |
| ARANDU MIRIM | 4 |
| ARUJA - SP | 0 |
| BANANAL | 32 |
| BARRA FUNDA | 0 |
| BOA VISTA | 176 |
| BOM RETIRO | 0 |
| BRASILANDIA | 0 |
| CACHOEIRINHA | 0 |
| CAMBIRELA | 35 |
| CAMBOINHAS | 52 |

*Figura 3. Portal que fornece relatórios completos de dados coletados pelas DSEIs*

A consulta por DSEI pode retornar a quantidade de indivíduos existentes em cada etnia, bem como sua localização e aldeia pertencente, conforme *imagem* abaixo.

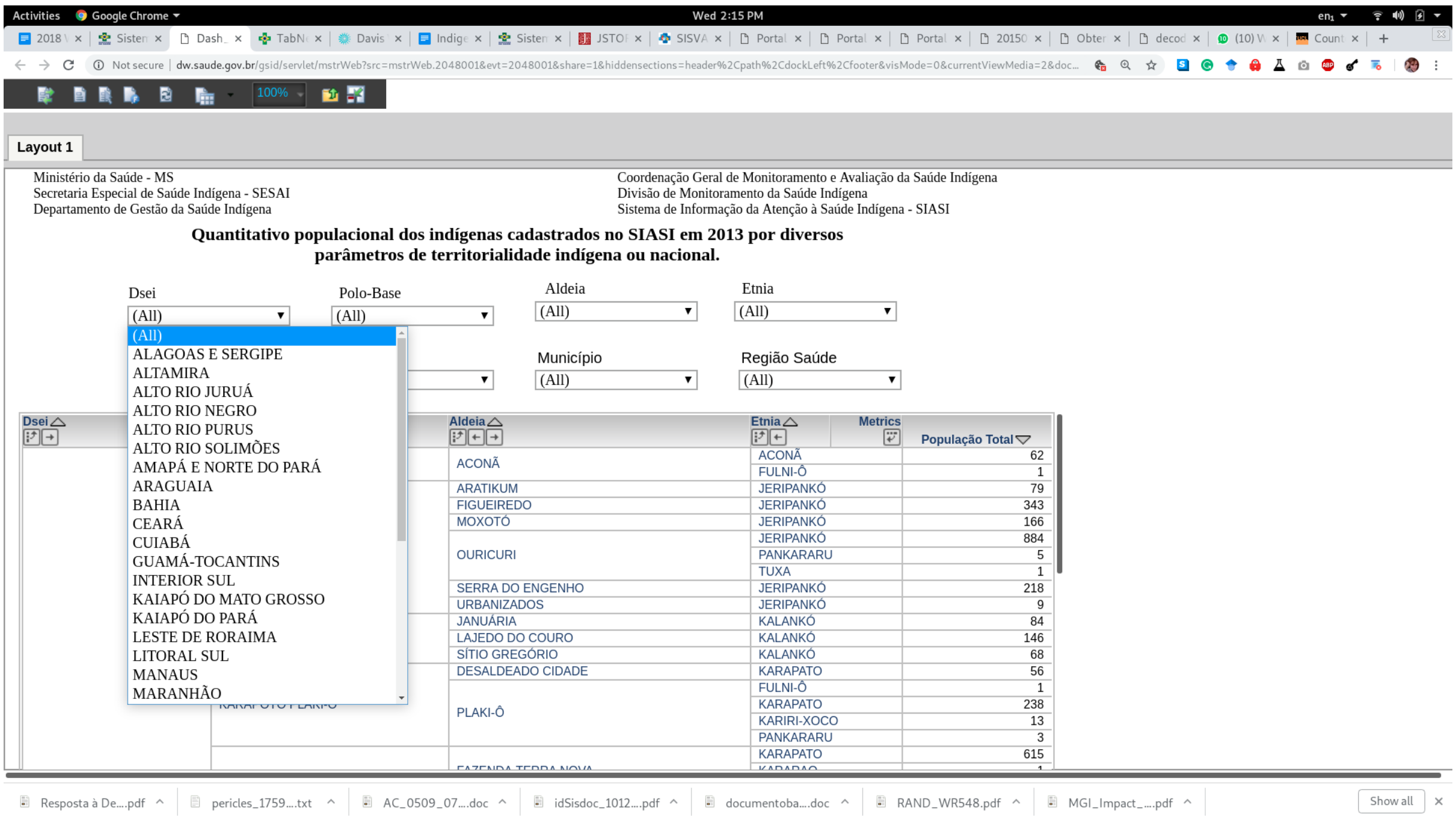

| Aldeia | Etnia | Metrics População Total |
|---|---|---|
| ACONÃ | ACONÃ | 62 |
| | FULNI-Ô | 1 |
| ARATIKUM | JERIPANKÓ | 79 |
| FIGUEIREDO | JERIPANKÓ | 343 |
| MOXOTÓ | JERIPANKÓ | 166 |
| OURICURI | JERIPANKÓ | 884 |
| | PANKARARU | 5 |
| | TUXA | 1 |
| SERRA DO ENGENHO | JERIPANKÓ | 218 |
| URBANIZADOS | JERIPANKÓ | 9 |
| JANUÁRIA | KALANKÓ | 84 |
| LAJEDO DO COURO | KALANKÓ | 146 |
| SÍTIO GREGÓRIO | KALANKÓ | 68 |
| DESALDEADO CIDADE | KARAPATO | 56 |
| PLAKI-Ô | FULNI-Ô | 1 |
| | KARAPATO | 238 |
| | KARIRI-XOCO | 13 |
| | PANKARARU | 3 |
| | KARAPATO | 615 |

*Figura 4. Imagem de consulta por DSEI*

É possível fazer o download de relatórios completos de 2013, em formato HTML ou em planilhas para excel. Assim como o sistema não apresenta atualização do http para o https, ou seja, é inseguro e permite a intermediação dos pacotes de informação trocados com o servidor, não fornece registro ou controle algum sobre a distribuição dessas informações.

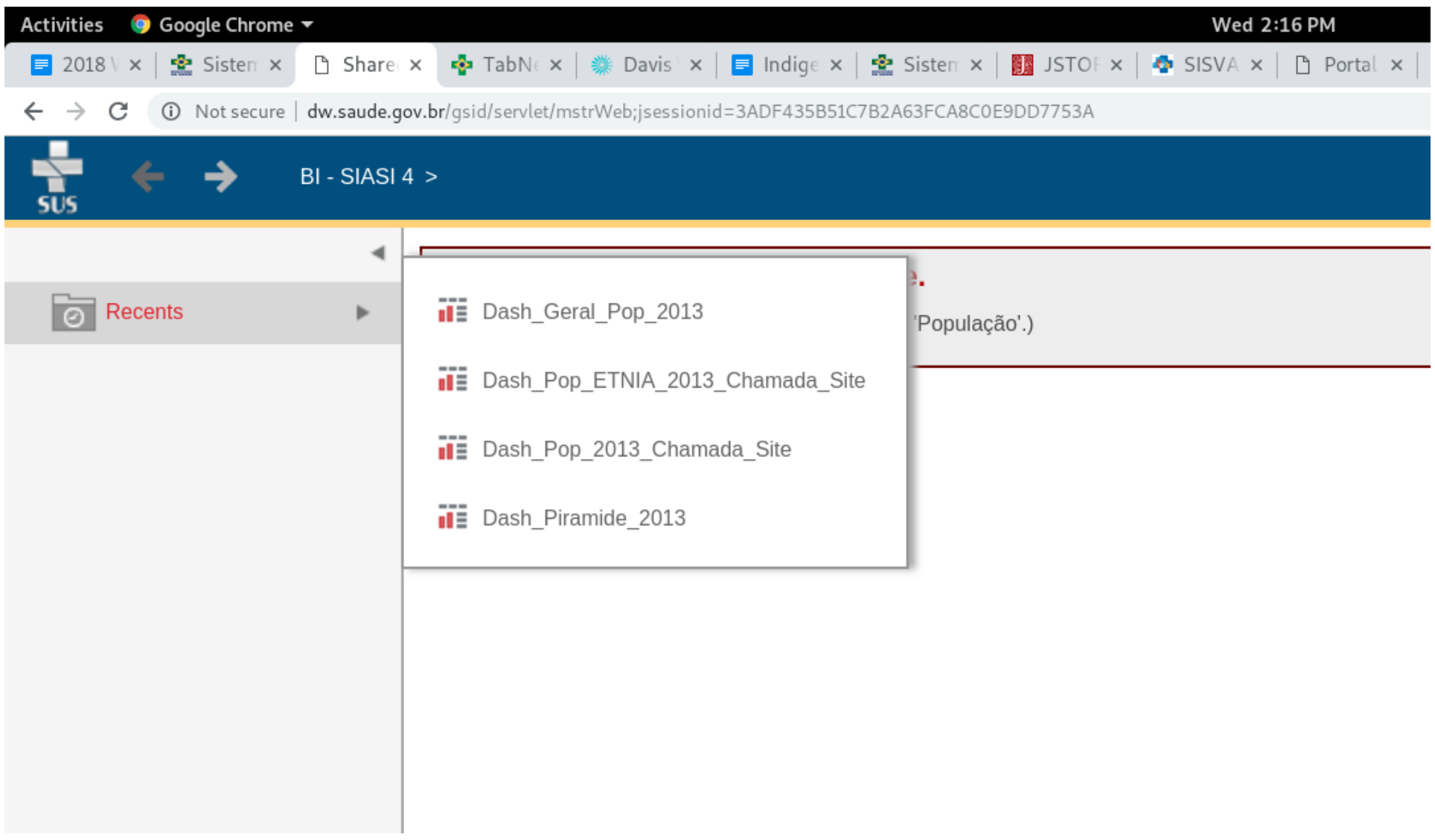

*Figura 5. Download de relatórios completos*

A exibição de tais dados é danosa às comunidades, pois a exposição de dados sobre a localização geográfica dos indígenas, com detalhamentos sobre a composição das aldeias ou acampamentos, é informação valiosa para o já denunciado extermínio de indígenas[7], prática que tem se intensificado no Brasil nos últimos anos, especialmente devido à disputas de terras.

## 2) Gestão dos dados indígenas: responsáveis

A gestão dos sistemas que sustentam a coleta, armazenamento, distribuição e integração desses dados se dá como consequência de uma legislação que vem sido modificada através de vários dispositivos, entre eles portarias, decretos e Leis aprovadas. Cada mudança na legislação acaba acarretando em modificações na coleta, tratamento e envio de dados. A migração de sistemas de dados também é um problema, assim como seu desligamento e arquivamento ou destruição dos dados antigos.

As mudanças de legislação também provocam modificações nas responsabilidades, o que se reflete na resposta à solicitação de informações realizada para a produção desse relatório (SIC nº 2489210), onde o órgão responsável atesta que a responsabilidade pela gestão dos dados é dividida entre o Datasus e o DGISI/DGESI/SESAI do Ministério da Saúde. Em termos de execução e manutenção dos sistemas que armazenam os dados, existe uma empresa contratada para realizar a manutenção dos sistemas por demanda. A empresa em questão é a CTIS, com sede em Brasília. Os termos do contrato não especificam detalhadamente os parâmetros da contratação. A página do Ministério que abriga o SIASI menciona a transmissão criptografada dos dados e também que "No campo da disseminação de dados o sistema está utilizando a estratégia de inteligência de negócio (Business Intelligence - BI) com a formação de um Data Warehouse (DW), potencializando a integração e análise de dados, além da integração com outros sistemas de informação de saúde".

[7] Fazendeiros formaram milícia para atacar índios no Mato Grosso do Sul, diz MPF. El País edicão digital. Acessado em 17 de novembro de 2017. Disponível em :https://brasil.elpais.com/brasil/2016/06/17/politica/1466195701_933817.html

Vale lembrar que, com frequência, tais dados são utilizados em conjunto com dados Sistema de Gestão do Programa Bolsa Família na Saúde (SISPBF) e do e-SUS/AB. Todos sistemas que contém informações que não deveriam ser publicadas. Tais sistemas e requisitos não[8] mencionam cuidados com a segurança na obtenção, acesso e transmissão e uso de tais informações. Apesar de o Ministério garantir que: "somente os profissionais da Secretaria Especial da Saúde Indígena (SESAI) podem acessar o SIASI Web", a exibição de dados em diversas instâncias na Web fragiliza as proteções que deveriam ser estendidas à essas comunidades.

Abaixo um exemplo de quais dados estão disponíveis no e-SUS, um sistema que concentra informações de saúde de pessoas que dependem do Sistema único de saúde. Tais informações são predominantemente sobre pessoas de populações vulneráveis, que fazem uso de seu "cartão do SUS".

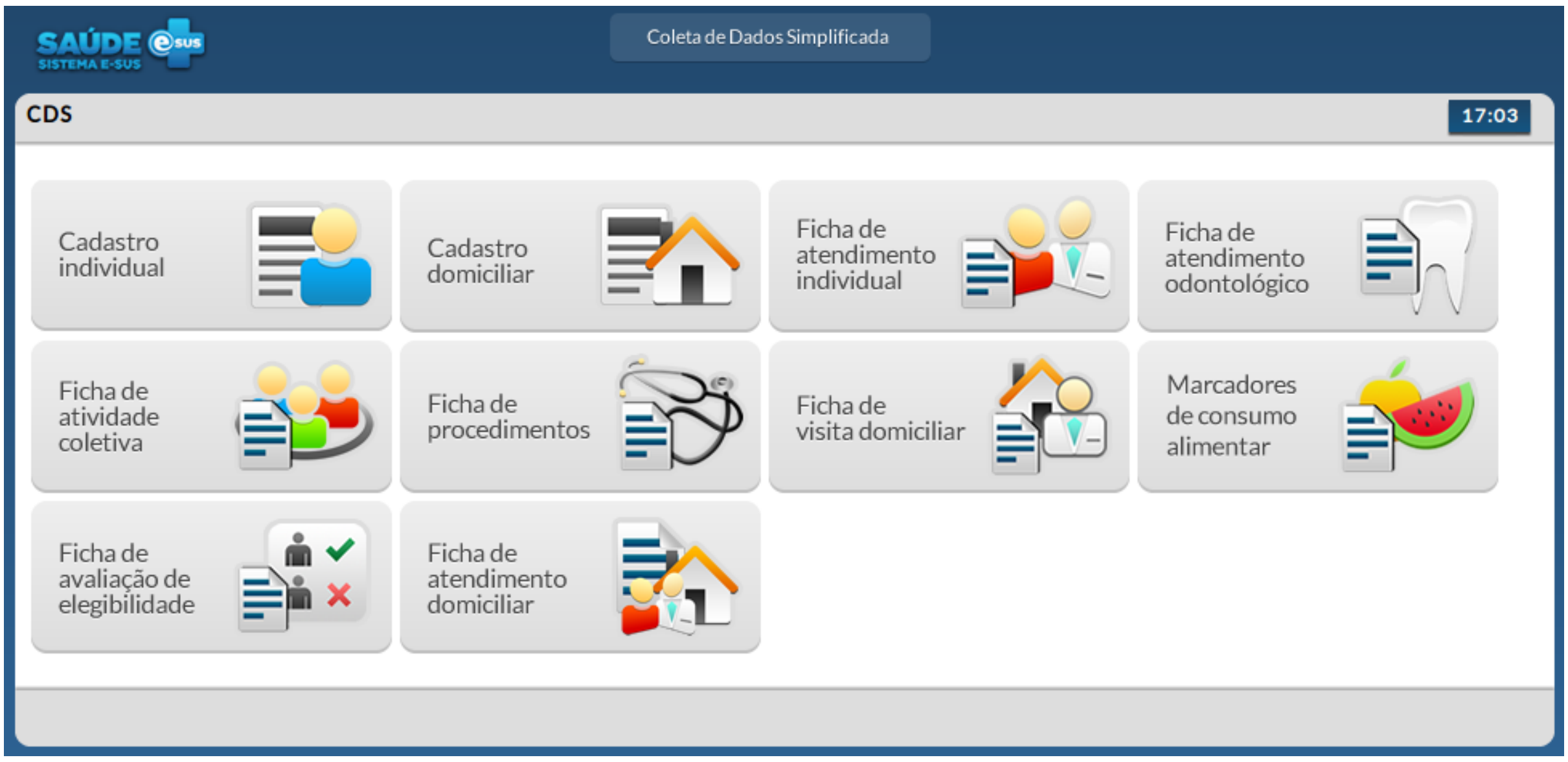

*Figura 6: Tela Inicial do Sistema e-SUS AB*

[8] Sistema e-SUS Atenção Básica Accessado em 12 de novembro de 2017. Disponível em http://dab.saude.gov.br/portaldab/esus/manualExportacaoV20/ManualExportacao_eSUSABv2.0.html#h.4e0ta940ebcu

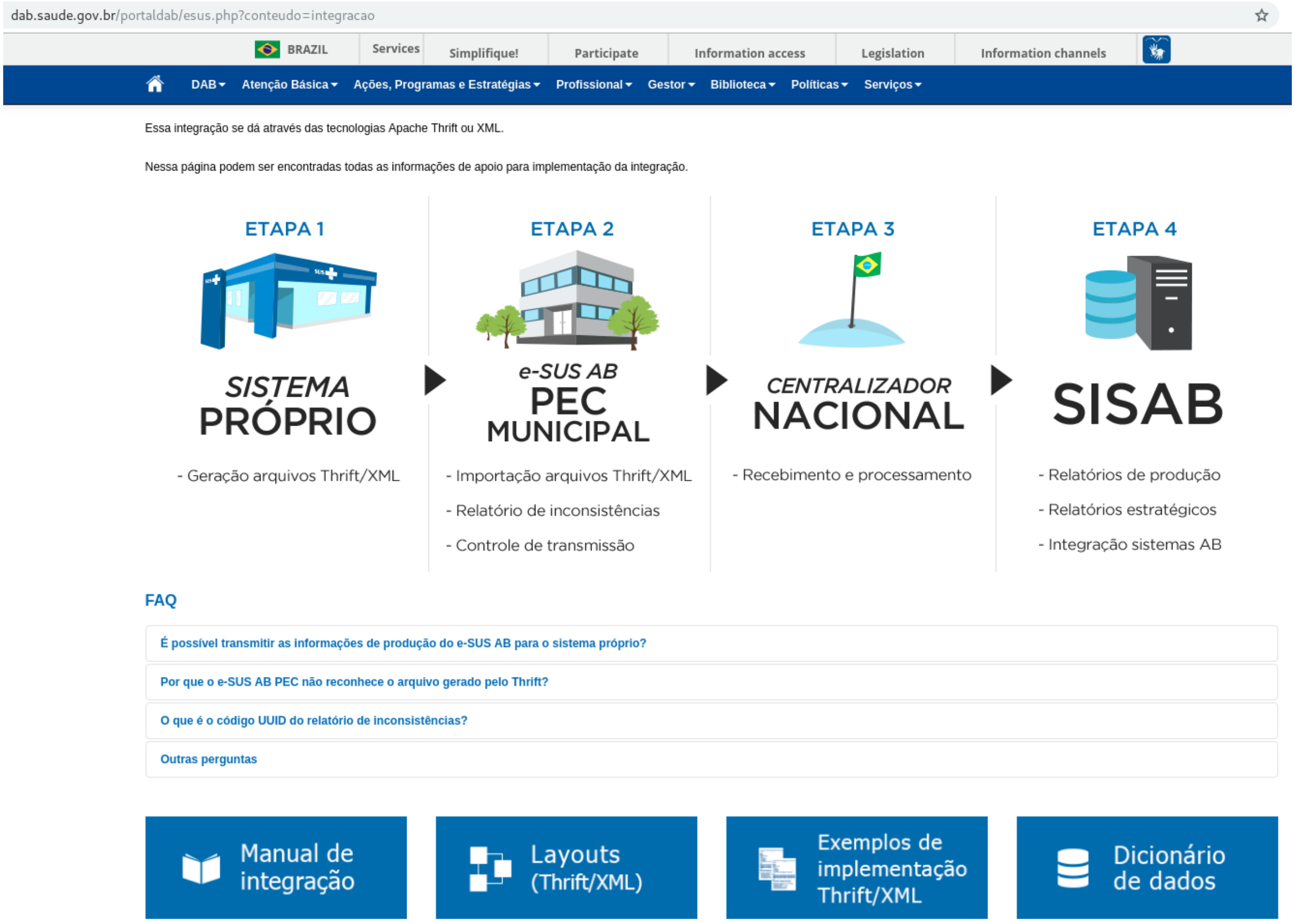

*Figura 7: Portal do Ministério da Saúde: não há preocupação com sigilo de dados*

Atualmente o DATASUS possui contrato com a CTIS, uma empresa com sede em Brasília, para desenvolvimento de software. Quanto ao custo, é necessário que seja especificado o período da informação desejada, já que a empresa atua conforme demanda, de acordo com as necessidades dos sistemas no período. É obscuro o esquema que determina quais são os responsáveis em caso de problemas enfrentados pelas populações em caso de vazamento desses dados.

## 3) Conclusão

Apesar das propostas de gestão participativa na organização da saúde indígena, disponíveis no documento que foi utilizado como base para a implementação[9] e aperfeiçoamento das plataformas de dados e sua gestão, não existem rastros de que o cumprimento de tais premissas tenha sido reforçado durante a implementação dos sistemas que digitalizam informações sobre as comunidades indígenas no Brasil. Muito pelo contrário, o desrespeito e descaso total com a privacidade, incluindo informações sobre a localização geográfica de grupos inteiros, por exemplo, ou de suas peculiaridades em relação à saúde, mostra que não existe uma preocupação com a participação das comunidades no modo como os dados são distribuídos e armazenados. Um dos destaques de Maria da Conceição de Sousa; João Henrique G. Scatena e Ricardo Ventura Santos; em seu paper “O Sistema de Informação da Atenção à Saúde Indígena (SIASI): criação, estrutura e funcionamento” tem como conclusão o seguinte trecho:

> “No caso do SIASI, vê-se uma situação na qual a coleta, o armazenamento e o processamento dos dados, apesar dos recursos de informatização disponíveis e dos problemas apontados, não vieram associados a uma ampliação, na escala necessária, da capacidade de análise e de emprego das informações em saúde para fins da execução e avaliação das ações.”

[9] DISTRITO SANITÁRIO ESPECIAL INDÍGENA: TERRITÓRIO DE PRODUÇÃO DE SAÚDE, PROTEÇÃO DA VIDA E VALORIZAÇÃO DAS TRADIÇÕES. Acessado em 17 de setembro de 2018/ Disponível em http://conselho.saude.gov.br/wsi/documentobase.doc

A escala, nesse caso, ou a ampliação dos sistemas que coletam dados provocam potencial monitoramento das comunidades indígenas. Dados de mortalidade indígena no Brasil indicam que a coleta de dados tem sido inócua para a proteção de tais comunidades, ainda que a coleta de dados tenha se intensificado nos últimos anos, inclusive tornando-se invasiva em termos de privacidade. O fato de que dados sobre acampamentos que ainda lutam pela demarcação de terras estejam expostos na web, juntamente com a descrição básica da população que nela está, por exemplo, demonstra que o entendimento dos gestores sobre a função dos dados, que deveria ser proteger e dar assistência à comunidades indígenas, foi desvirtuada pela própria coleta e distribuição de tais dados. Há uma preocupação em coletar cada vez mais dados pessoais, reforçando o controle de corpos e a destruição da cultura desses povos.

A digitalização pode colaborar na continuação do genocídio da população indígena. Mais do que nunca as tecnologias de análise e coleta de dados precisam ser postas em discussão, uma vez que reforçam as separações étnicas e a discriminação de brasileiros que têm direito à sua terra e à sua cultura. Ante a vasta disponibilidade de dados estruturados e séries históricas que permitem a análise e correlação destes dados com outros eventos ocorridos com essas populações, não é difícil concluir: já se tem suficientes evidências de que a degradação de terras, o apagamento de culturas originais e o assassinato de lideranças indígenas são os principais representantes das ameaças à existência desses povos. Ainda, dados sobre as florestas e recursos naturais que poderiam ser utilizados como indicadores sobre a existência e conservação dessas culturas podem e devem ser analisados no lugar de dados pessoais minuciosos, ou ainda dados sobre a localização dessas populações exibidos ao público. Conforme relatório do ISA de 2000:

> “As prioridades ambientais para uma política de atenção à saúde dos povos indígenas devem contemplar a preservação das fontes de água limpa, construção de poços ou captação à distância nas comunidades que não dispõem de água potável; a construção de sistema de

> esgotamento sanitário e destinação final do lixo nas comunidades mais populosas; a reposição de espécies utilizadas pela medicina tradicional; e o controle de poluição de nascentes e cursos d'água situados acima das terras indígenas.[10]"

Os sistemas de coleta de dados de indígenas no Brasil precisam ser colocados em discussão. A função dessas bases de dados precisa ser desafiada, juntamente com a sua existência. Finalmente, não há um dispositivo único que trate dos dados de indígenas. O Estatuto do Índio[11], a Lei de nº 6.001 de 1973 - sequer cita a palavra "dado", o que deixa claro que é necessária uma reforma legislativa sobre o tema. A inovação tecnológica precisa ser abrangente, inclusiva e diversa, trazendo benefícios para todos. Infelizmente, tais populações têm servido como laboratórios de ensaios para experimentos de acompanhamento de dados que envolvem a exposição de comunidades inteiras.

**Referências**

---

[10] Informe Indígena ISA https://acervo.socioambiental.org/sites/default/files/documents/F2D00064.pdf
[11] LEI Nº 6.001, DE 19 DE DEZEMBRO DE 1973. http://www.planalto.gov.br/ccivil_03/Leis/L6001.htm